\documentstyle[twocolumn,epsf]{IEEEtran}
\def\BibTeX{{\rm B\kern-.05em{\sc i\kern-.025em b}\kern-.08em
            T\kern-.1667em\lower.7ex\hbox{E}\kern-.125emX}}

\newcommand{\up}{\uparrow}
\newcommand{\dwn}{\downarrow}
\newcommand{\ev}[1]{\langle #1 \rangle}

\newcommand{\lesssim}{\ \raise.3ex\hbox{$<$}\kern-0.8em\lower.7ex\hbox{$\sim$}~ }
\newcommand{\gesim}{\ \raise.3ex\hbox{$>$}\kern-0.8em\lower.7ex\hbox{$\sim$}~ }

\input epsf

\begin{document}
\title{Superfluid/ferromagnet/superfluid-junction and $\pi$-phase in a superfluid Fermi gas at finite temperatures}
\author{\vskip 1em
{\Large Takashi Kashimura, S. Tsuchiya, and Y. Ohashi}\thanks{T. Kashimura is with Department of Physics, Keio University, Yokohama, Japan. (E-mail:t.kashimura@a3.keio.jp) S. Tsuchiya is with Department of Physics, Tokyo University of Science, Tokyo, Japan, and CREST(JST), Saitama, Japan. Y. Ohashi is with Department of Physics, Keio University and CREST(JST).}
\vskip 1em
}
%\markboth{{\small COE RA Educational Program, KO-1, 1st April, 2007}}{Murray and Balemi: Using the style file IEEEtran.cls}
\maketitle

%%%%%%%%%%%%%%%%%%%%%%%%%%%%%%%%%%%%%  Abstract  %%%%%%%%%%%%%%%%%%%%%%%%%%%%%%

\begin{abstract}
We investigate the stability of $\pi$-phase in a polarized superfluid Fermi gas ($N_\uparrow>N_\downarrow$, where $N_\sigma$ is the number of atoms in the hyperfine state described by pseudospin-$\sigma$). In our previous paper [T. Kashimura, S. Tsuchiya, and Y. Ohashi, Phys. Rev. A \textbf{82}, 033617 (2010)],  we showed that excess atoms ($\Delta N=N_\uparrow-N_\downarrow$) localized around a potential barrier embedded in the system induces the $\pi$-phase at $T=0$, where the phase of superfluid order parameter differ by $\pi$ across the junction. In this paper, we extend our previous work to include temperature effects within the mean-field theory. We show that the $\pi$-phase is stable even at finite temperatures, although transition from the $\pi$-phase to 0-phase eventually occurs at a certain temperature. Our results indicate that the $\pi$-phase is experimentally accessible in cold Fermi gases.
\end{abstract}

%%%%%%%%%%%%%%%%%%%%%%%%%%%%%%%%%%%%%%%%%%%%%%%%%%%%%%%%%%%%%%%%%%%%%%%%%%%%%%


%%%%%%%%%%%%%%%%%%%%%%%%%%%%%%%%%%%%%%%%%%%%%%%%%%%%%%%%%%%%%%%%%%%%%%%%%%%%%%

\section{Introduction} \label{sec1}

\PARstart{T}{he} recently realized superfluid $^{40}$K \cite{Regal} and $^6$Li \cite{Zwierlein0,Kinast,Bartenstein} Fermi gases are expected as useful `quantum simulators' for metallic superconductivity. This is because, while the background physics of superfluid Fermi gases is the same as superconductivity, the former system has the unique property that various physical parameters can be experimentally tuned. In particular, We can study superfluid properties of this system from the weak-coupling regime to the strong-coupling regime in a unified manner \cite{Ohashi} (which is sometimes referred to the BCS-BEC crossover in the literature), by adjusting a tunable pairing interaction associated with a Feshbach resonance \cite{Timmermans}. Thus, this system is expected to be useful for the study of strongly correlated high-$T_{\rm c}$ cuprates. Indeed, using this advantage, the so-called pseudogap phenomenon, which has been extensively discussed in the underdoped regime of high-$T_{\rm c}$ cuprates \cite{Lee,Fischer}, was recently observed in the BCS-BEC crossover regime of $^{40}$K Fermi gases \cite{Stewart,Gaebler}. 
\par
In our previous paper  \cite{Kashimura}, as another application of superfluid Fermi gases to condensed matter physics, we theoretically proposed an idea to simulate a superfluid/ferromagnet/superfluid (SFS)-junction. By numerically solving the mean-field Bogoliubov-de Gennes theory in real space at $T=0$, we showed that, when we put a {\it nonmagnetic} potential barrier in a polarized superfluid Fermi gas ($N_\uparrow>N_\downarrow$, where $N_\sigma$ is the number of Fermi atoms in the atomic hyperfine state described by pseudospin-$\sigma$), it is magnetized in the sense that some of excess $\uparrow$-spin atoms are localized around it. The superfluid Fermi gas is then separated by this {\it pseudo-ferromagnetic junction}, the structure of which is similar to a superconductor/ferromagnet/superconductor (SC/F/SC)-junction in metallic superconductors  \cite{Buzdin}. We also showed that the SFS-junction really works as a ferromagnetic junction, because the so-called $\pi$-phase (which is a typical phenomenon realized in the SC/F/SC-junction) is stably realized, where the superfluid order parameter changes its sign across the junction. 
\par
In this paper, we extend our previous work at $T=0$  \cite{Kashimura} to the case of finite temperatures. Since experiments are always done at finite temperatures, this extension is important to clarify to what extent the SFS-junction, as well as the $\pi$-phase, are stable against thermal effects. In a two-dimensional attractive Fermi Hubbard model with population imbalance, we examine the superfluid order parameter, as well as local population imbalance (or magnetization), by numerically solving the Bogoliubov-de Gennes equations. We show that the $\pi$-phase stably exists even at finite temperatures, so that one can experimentally access this interesting phase.
\par
%%%%%%%%%%%%%%%%%%%%%%%%%%%%%%%%%%%%%%%%%%%%%%%%%%%%%%%%%%%%%
%%%%%%%%%%%%%%%%%%%%%%%%%%%%%%%%%%%%%%%%%%%%%%%%%%%%%%%%%%%%%%%%%%%%%%%%%%%%
\begin{figure}[ttt]
\centering
\epsfxsize=7.5cm
\epsfbox{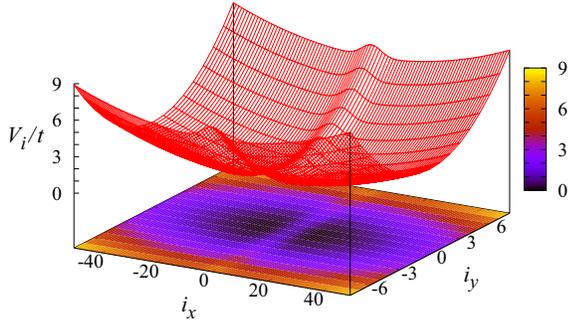}\\
\caption{Spatial variation of model double-well potential $V_i$ used in this paper. We take $V_{0x}^t=0.001t$, $V_{0y}^t=0.1t$, $V_0^b=0.1t$, and $\ell=5$ (where the lattice constant is taken to be unity). These parameters are used throughout this paper.}
\label{fig1}
\end{figure}%
%%%%%%%%%%%%%%%%%%%%%%%%%%%%%%%%%%%%%%%%%%%%%%%%%%%%%%%%%%%%%%%%%%%%%%%%%%%%%%
\section{Formalism} \label{sec2}

We consider a two-component Fermi gas decribed by pseudospin $\sigma=\uparrow,\downarrow$. To investigate the stability of $\pi$-phase induced by a SFS-junction in a simple manner, we treat a polarized Fermi gas described by the two-dimensional Hubbard Hamiltonian,
\begin{eqnarray} \label{H} 
H=-t\sum_{\langle i,j \rangle, \sigma} 
\left[ \hat{c}_{i,\sigma}^\dagger \hat{c}_{i,\sigma} + h.c. \right] 
-U \sum_i \hat{n}_{i,\uparrow} \hat{n}_{i,\downarrow} 
\nonumber \\
- \sum_{i,\sigma} \left[ \mu_\sigma - V_i \right] \hat{n}_{i,\sigma}.
\end{eqnarray}
Here, $\hat{c}^\dagger_{i,\sigma}$ is the creation operator of a Fermi atom with pseudospin $\sigma=\uparrow,\downarrow$ at the $i$-th site. $-t$ is a nearest-neighbour hopping energy, and the summation $\langle i,j \rangle$ is taken over the nearest-neighbour pairs. $-U (<0)$ is an on-site pairing interaction, and $\hat{n}_{i,\sigma} \equiv \hat{c}_{i,\sigma}^\dagger \hat{c}_{i,\sigma}$ is the number operator at the $i$-th site. Since we consider a polarized system in this paper, the chemical potential $\mu_\sigma$ depends on pseudospin $\sigma$. $V_i$ is a double-well potential, given by
\begin{eqnarray} \label{V} 
V_i = V_{0x}^t i_x^2 + V_{0y}^t i_y^2 + V_0^b \exp{\left[-(i_x/\ell)^2\right]},
\end{eqnarray}
where $i=(i_x,i_y)$ is the spatial position of the $i$-th lattice site. The first two terms describe a cigar trap (or anitotropic harmonic trap), and the last term gives a potential barrier around $i_x=0$. $\ell$ is the width of the barrier potential. We explicitly show the spatial variation of Eq. (\ref{V}) in Fig. \ref{fig1}. 
\par

In the mean-field approximation, the model Hamiltonian in Eq. (\ref{H}) reduces to
\begin{eqnarray}
H_{\rm MF} 
&=& 
-t \sum_{\langle i,j \rangle,\sigma} 
\left[
\hat{c}^\dagger_{i,\sigma} \hat{c}_{j,\sigma}+ {\rm h.c.} 
\right] 
\nonumber
\\
&-&
\sum_i
\left[
\Delta_i\hat{c}_{i,\up}^\dagger \hat{c}_{i,\dwn}^\dagger 
+ \Delta_i^*\hat{c}_{i,\dwn} \hat{c}_{i,\up}
\right]
\nonumber
\\
&+&
\sum_{i,\sigma} 
\left[
V_i-\mu_\sigma-U \ev{\hat{n}_{i,-\sigma}} 
\right]
\hat{n}_{i,\sigma}
\nonumber
\\
&+&
\sum_{i} 
\left[
\frac{\Delta_i^2}{U} + U \ev{\hat{n}_{i,\up}} 
\ev{\hat{n}_{i,\dwn}} 
\right],
\label{MFH}
\end{eqnarray}
where $\Delta_i=U\langle \hat{c}_{i,\downarrow}\hat{c}_{i,\uparrow}\rangle$ is the superfluid order parameter, which is taken to be real in this paper. Noting that Eq. (\ref{MFH}) is a bilinear form with respect to the creation and annihilation operators, as usual, one can conveniently diagonalize it by the Bogoliubov transformation \cite{Kashimura},
\begin{eqnarray}
\left(
\begin{array}{c}
{\hat c}_{1,\uparrow}\\
{\hat c}_{2,\uparrow}\\
:\\
{\hat c}_{M,\uparrow}\\
{\hat c}_{1,\downarrow}^\dagger\\
{\hat c}_{2,\downarrow}^\dagger\\
:\\
{\hat c}_{M,\downarrow}^\dagger\\
\end{array}
\right)
=
{\hat W}
\left(
\begin{array}{c}
{\hat \alpha}_{1,\uparrow}\\
{\hat \alpha}_{2,\uparrow}\\
:\\
{\hat \alpha}_{M,\uparrow}\\
{\hat \alpha}_{1,\downarrow}^\dagger\\
{\hat \alpha}_{2,\downarrow}^\dagger\\
:\\
{\hat \alpha}_{M,\downarrow}^\dagger\\
\end{array}
\right),
\label{eq.BB}
\end{eqnarray}
Here, ${\hat W}$ is a $2M\times 2M$-orthogonal matrix, where $M$ is the total number of lattice sites. We briefly note that the determination of ${\hat W}$ corresponds to solving the Bogoliubov-de Gennes equations in the case of uniform Fermi superfluid \cite{deGennes}. 
\par
The diagonalized mean-field Hamiltonian has the form \cite{Kashimura}
\begin{eqnarray}
H_{\rm MF}
=
\sum_{j=1}^M E_{j,\sigma} \hat{\alpha}_{j,\sigma}^\dagger \hat{\alpha}_{j,\sigma} 
+
E_{G0}
\label{DH}
\end{eqnarray}
where $E_{j,\sigma}$ is the Bogoliubov single-particle excitation spectrum, and \begin{eqnarray}
E_{G0}
&=&
\sum_{i=1}^M 
\Bigl[
V_i -\mu_\downarrow- U\langle\hat{n}_{i,\up}\rangle
\nonumber
\\
&+&\frac{\Delta_i^2}{U}
+ U \langle \hat{n}_{i,\up}\rangle \langle\hat{n}_{i,\dwn}\rangle 
 -E_{i,\downarrow}
\Bigr].
\label{EG0}
\end{eqnarray}
\par
The superfluid order parameter $\Delta_i$ and the particle density $\langle {\hat n}_{i,\sigma}\rangle$ are given by, respectively,
\begin{eqnarray}
\Delta_i 
= 
&U& \sum_{j=1}^M
\Bigl[
W_{i,j} W_{M+i,j} f(E_{j,\up} )
\nonumber
\\
&+& W_{i,M+j} W_{M+i,M+j} f(-E_{M+1-j,\downarrow})
\Bigr],
\label{GAP}
\end{eqnarray}
\begin{eqnarray}
\ev{\hat{n}_{i,\up}} 
= 
\sum_{j=1}^M 
\Bigl[
W_{i,j}^2 f(E_{j,\up} )
+
 W_{i,M+j}^2 f(-E_{M+1-j,\downarrow})
\Bigr],
\end{eqnarray}
\begin{eqnarray} 
\ev{\hat{n}_{i,\dwn}} 
&=&
\sum_{j=1}^M 
\Bigl[
W_{M+i,j}^2 f(-E_{j,\up} )
\nonumber
\\
&+& W_{M+i,M+j}^2 f(E_{M+1-j,\downarrow}) 
\Bigr].
\label{DENSITY}
\end{eqnarray}
Here, $f(E)=1/(e^{E/T}+1)$ is the Fermi distribution function (where we take $k_{\rm B}=1$). At $T=0$, the Fermi distribution function reduces to the step function as $f(E)\to \Theta(-E)$, so that Eqs. (\ref{GAP})-(\ref{DENSITY}) reproduce Eqs.(6) and (7) in Ref. \cite{Kashimura}.
\par
In our numerical calculations, we self-consistently determine $\Delta_i$, $\langle \hat{n}_{i,\sigma} \rangle$, and $\mu_\sigma$, for a given parameter set ($N_\up, N_\dwn, U, V_{x0}, V_{y0}, V_0^b, \ell$). We energetically compare the $\pi$-phase solution with the $0$-phase solution (where the superfluid order parameter has the same sign bwteen the left and right sides of the SFS-junction). The free energy for fixed particle numbers $N_\uparrow$ and $N_\downarrow$ is given by
\par
\begin{eqnarray} \label{E_G}
F = -T\sum_{j,\sigma} \log [1+e^{-E_{j,\sigma}/T}]+ \sum_\sigma \mu_\sigma N_\sigma +E_{G0}.
\end{eqnarray}
%%%%%%%%%%%%%%%%%%%%%%%%%%%%%%%%%%%%%%%%%%%%%%%%%%%%%%%%%%%%
\par
\section{Temperature dependence of SFS junction and $\pi$ phase} \label{sec3}

%%%%%%%%%%%%%%%%%%%%%%%%%%%%%%%%%%%%%%%%%%%%%%%%%%%%%%%%%%%%%%%%%%%%%%%%%%%%%%
\begin{figure}[!t]
\centering
\epsfxsize=7.5cm
\epsfbox{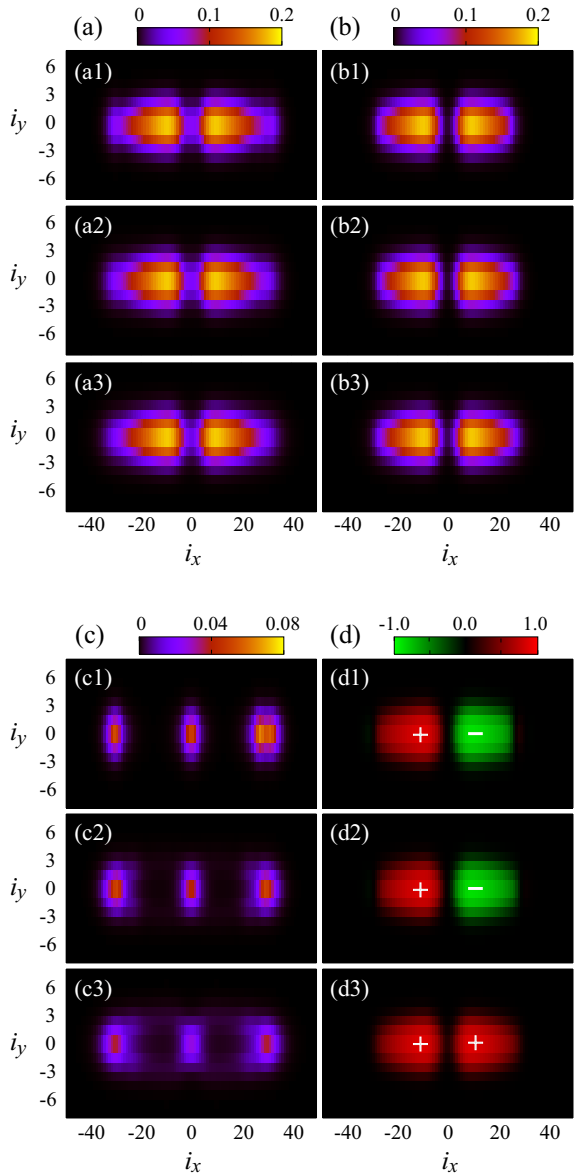}\\
\caption{Calculated superfluid state in the double-well potential $V_i$ at finite temperatures, in the presence of population imbalance ($N_\up=32>N_\dwn=28$). The upper, middle, and lower panels show the results for $T=0$, $0.15T_{\rm c}$, and $0.25T_{\rm c}$, respectively (where $T_{\rm c}=0.4t$ in this calculation). Panels (a) and (b) show the particle densities $\langle {\hat n}_{i,\sigma}\rangle$ of $\uparrow$-spin atoms and $\downarrow$-spin atoms, respectively. Panels (c) show the local polarization $s^z_i=\langle {\hat n}_{i,\uparrow}\rangle- \langle {\hat n}_{i,\downarrow}\rangle$. Panels (d) show the spatial variation of the superfluid order parameter $\Delta_i$. We take $U/t=4$.
}
\label{fig2}
\end{figure}%
%%%%%%%%%%%%%%%%%%%%%%%%%%%%%%%%%%%%%%%%%%%%%%%%%%%%%%%%%%%%%%%%%%%%%%%%%%%%%%%

\begin{figure}[ttt]
\centering
\epsfxsize=7cm
\epsfbox{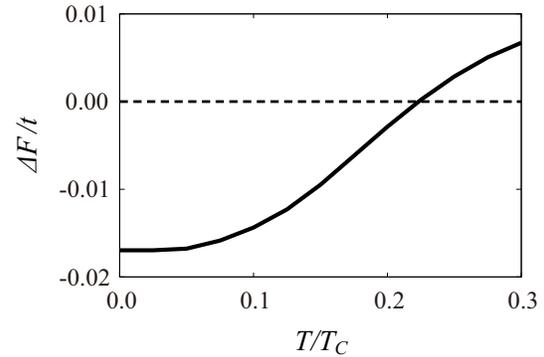}\\
\caption{Calculated difference of free energy $\Delta F = F_{\pi} - F_{0}$ between the $\pi$-phase ($F_\pi$) and 0-phase ($F_0$), as a function of temperature $T$. 
}
\label{fig3}
\end{figure}

Figure  \ref{fig2} shows the temperature dependence of supefluid state around the SFS-junction. At $T=0$ shown in the upper panels, one finds that the SFS-junction and $\pi$ phase are realized. That is, panel (c1) shows that some of excess atoms are localized around the central barrier potential ($i_x\sim 0$). Because of this local {\it magnetization} around the barrier, the superfluid order parameter $\Delta_i$ changes its sign across the junction, as shown in panel (d1), which is just the $\pi$-phase. We note that, although we also obtain the 0-phase solution, it is energetically unfavorable, as shown in Fig. \ref{fig3}. Since the $\pi$-phase is known as a characteristic phenomenon in a ferromagnetic junction, we find that the localized excess atoms with pseudospin-$\uparrow$ really behave like magnetic spins in the present system.
\par
The $\pi$-phase can still stably exist at finite temperatures, as shown in Fig. \ref{fig2}(d2) and Fig. \ref{fig3}. However, we also find in Fig. \ref{fig2}(c2) that the localized excess atoms gradually spread out due to thermal excitations, leading to the weakening of the `magnetization' of the SFS-junction. As a result, the 0-phase become more stable than the $\pi$-phase when $T\ge 0.22T_{\rm c}$, as shown in Fig. \ref{fig3}. At $T=0.25T_{\rm c}>0.22T_{\rm c}$, although we can still see finite magnetization of the central potential barrier in Fig. \ref{fig2}(c3), the superfluid order parameter no longer changes its sign across the junction (0-phase), as shown in Fig. \ref{fig2}(d3). 

%%%%%%%%%%%%%%%%%%%%%%%%%%%%%%%%%%%%%%%%%%%%%%%%%%%%%%%%%%%%%%%%%%%%%%%%%%%%%%
%%%%%%%%%%%%%%%%%%%%%%%%%%%%%%%%%%%%%%%%%%%%%%%%%%%%%%%%%%%%%%%%%%%%%%%%%%%%%%

\section{Summary}\label{sec4}
\par
To summarize, we have investigated the stability of $\pi$-phase in a polarized superfluid Fermi gas at finite temperatures. In a model two-dimensional attractive Fermi Hubbard model with a double-well potential, we showed that the SFS-junction and $\pi$ phase can stably exist at finite temperatures. Our results indicate that the SFS-junction and $\pi$-phase is experimentally accessible.
\par
In this paper, we have treated the model Fermi gas within the mean-field theory. In this regard, we note that pairing fluctuations, which are ignored in the present mean-field theory, are known to be crucial for ultracold Fermi gases, especially near the superfluid phase transition temperature $T_{\rm c}$. However, since the $\pi$-phase is stable only far below $T_{\rm c}$ ($T\lesssim 0.22T_{\rm c}\ll T_{\rm c}$), where superfluid fluctuations are almost suppressed by the superfluid excitation gap, we expect that our results obtained in this paper is not drastically altered when pairing fluctuations are included. On the other hand, since the SFS-junction is still stable at $T=0.22T_{\rm c}$, one may need to include this strong-coupling effect in clarifying the temperature at which the SFS-junction disappears, which remains as a future problem.

%%%%%%%%%%%%%%%%%%%%%%%%%%%%%
\subsection*{Acknowledgments}
This work was supported in part by a Grant in Aid for the 21st century Center of Excellence for Optical and Electronic Device Technology for Access Network from the Ministry of Education, Culture, Sports, Science, and Technology in Japan.

%by Global COE Program "High-Level Global Cooperation for Leading-Edge Platform on Access Space (C12)."

%%%%%%%%%%%%%%%%%%%%%%%%%

%%%%%%%%%%%%%%%%%%%%%%%%%

%%%%%%%%%%%%%%%%%%%%%

\begin{thebibliography}{99}
\bibitem{Regal} C. A. Regal, M. Greiner, and D. S. Jin, Phys. Rev. Lett. \textbf{92}, 040403 (2004).
\bibitem{Zwierlein0} M. W. Zwierlein, C. A. Stan, C. H. Schunck, S. M. F. Raupach, A. J. Kerman, and W. Ketterle, Phys. Rev. Lett \textbf{92}, 120403 (2004).
\bibitem{Kinast} J. Kinast, S. L. Hemmer, M. E. Gehm, A. Turlapov, and J. E. Thomas, Phys. Rev. Lett. \textbf{92}, 150402 (2004).
\bibitem{Bartenstein} M. Bartenstein, A. Altmeyer, S. Riedl, S. Jochim, C. Chin, J. Hecker Denschlag, and R. Grimm, Phys. Rev. Lett \textbf{92}, 203201 (2004).
\bibitem{Ohashi}Y. Ohashi and A. Griffin, Phys. Rev. Lett. \textbf{89}, 130402 (2002).
\bibitem{Timmermans}E. Timmermans, K. Furuya, P. W. Milonni, and A. K. Kerman, Phys. Lett. A \textbf{285}, 228 (2001).
\bibitem{Lee} P. Lee, N. Nagaosa, and X. Wen, Rev. Mod. Phys. \textbf{78}, 17 (2006).
\bibitem{Fischer} O. Fischer, M. Kugler, I. Maggio-Aprile, C. Berthod, and C. Renner, Rev. Mod. Phys. \textbf{79}, 353 (2007).
\bibitem{Stewart}J. T. Stewart, C. A. Regal, and D. S. Jin, Nature (London) \textbf{454}, 744 (2008).
\bibitem{Gaebler}J. P. Gaebler, J. T. Stewart, T. E. Drake, D. S. Jin, A. Perali, P. Pieri, and G. C. Strinati, Nature Phys. \textbf{6}, 569 (2010).
\bibitem{Kashimura} T. Kashimura, S. Tsuchiya, and Y. Ohashi, Phys. Rev. A \textbf{82}, 033617 (2010).
\bibitem{Buzdin} A. I. Buzdin, Rev. Mod. Phys. \textbf{77}, 935 (2005).
\bibitem{deGennes} See, for example, de Gennes, {\it Superconductivity of Metals and Alloys} (Addison-Wesley, New York, 1989).

\bibitem{Zwierlein} M. W. Zwierlein, A. Schirotzek, C. H. Schunck, and W. Ketterle, Science \textbf{311}, 492 (2006).
\bibitem{Partridge} G. B. Partridge, W. Li, R. I. Kamar, Y.-A. Liao, and R. G. Hulet, Science \textbf{311}, 503 (2006).

\end{thebibliography}
\end{document}